\documentclass[conference]{IEEEtran}

\usepackage{graphics} 
\usepackage{epsfig} 
\usepackage{mathptmx} 
\usepackage{times} 
\usepackage{amsmath} 
\usepackage{amssymb}  
\usepackage{stmaryrd}
\usepackage{color}

\usepackage{cite}

\ifCLASSINFOpdf
\else
\fi
\usepackage{dblfloatfix}
\usepackage{url}

\usepackage[scriptsize,tight]{subfigure}

\begin{document}
\title{The Velocity of the Propagating Wave\\ for General Coupled Scalar Systems}

\author{\IEEEauthorblockN{Rafah El-Khatib and Nicolas Macris}
\IEEEauthorblockA{LTHC, EPFL, Lausanne, Switzerland}
Emails: \{rafah.el-khatib,nicolas.macris\}@epfl.ch}


%


\maketitle

\begin{abstract}
We consider spatially coupled systems governed by a set of scalar density evolution equations. Such equations track the behavior
of message-passing algorithms used, for example, in coding, sparse sensing, or constraint-satisfaction problems. 
Assuming that the ``profile" describing the average state of the algorithm exhibits a solitonic wave-like behavior after initial transient iterations, we derive a formula for the propagation velocity of the wave. We illustrate the formula with two applications, namely Generalized LDPC codes and compressive sensing.
\end{abstract}

\IEEEpeerreviewmaketitle

\section{Introduction}
Spatial coupling is a graph construction that was first introduced for coding on Low-Density Parity-Check (LDPC) codes by Felstrom and Zigangirov 
\cite{Zigangirov}. Spatially coupled systems have been shown to exhibit excellent performance under low complexity message-passing 
algorithms. Due to this attractive property, they have been extensively studied in different frameworks, such as 
coding \cite{lentmaier2010iterative}, \cite{KRUUniv} (a review with applications in the broader context of communications is found in \cite{KRUUniv}), compressive sensing \cite{donoho2012information}, \cite{kudekar2010effect}, 
\cite{krzakala2012statistical}, statistical physics and constraint-satisfaction 
\cite{HassaniMacrisUrbanke}, \cite{hassani2010itw}, \cite{hassani2013threshold}, \cite{Achlioptas-et-al}. 

To assess the performance of a message-passing algorithm one analyzes the evolution of the messages exchanged during the algorithm as the number of iterations increases. This can be expressed as a set of scalar recursive equations (see Equ. \eqref{eqn:DEuncoupledDisc} for uncoupled systems and Equ. \eqref{eqn:DEcoupledDisc} for coupled systems). In coding these are the density evolution equations, and in compressive sensing these are the state evolution equations.

A spatially coupled system is obtained from the underlying ``single" (uncoupled) one by taking $2L+1$ copies of it and connecting every $W$ consecutive single systems by means of a ``coupling window".
For scalar systems, the average behavior of the system at position $i\in\{-L,\dots,L\}$ of the coupling axis and at iteration $t\in\mathbb{N}$ of the message-passing algorithm is described by a single scalar $x_i^{(t)}$. Therefore, the evolution of the coupled system can be analyzed by tracking the vector $\mathbf{x}^{(t)}=\{x_{-L}^{(t)},\dots,x_{L}^{(t)}\}$, which we call the ``profile", as $t$ increases.

\begin{figure}
\centering
\includegraphics[scale=0.27]{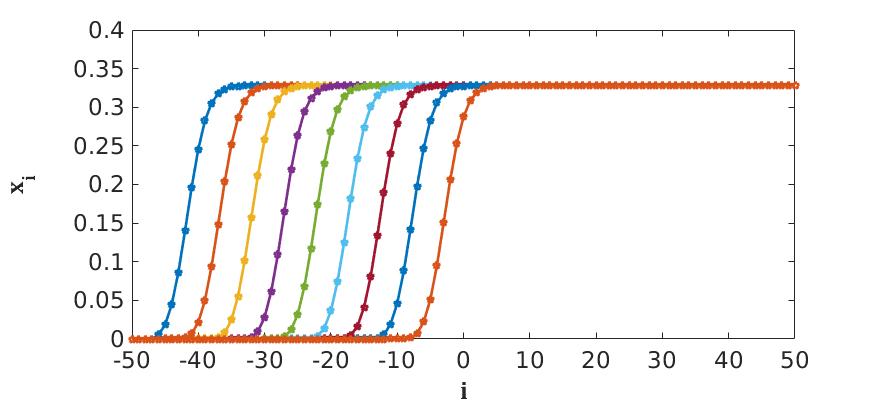}
\caption{The profile $\mathbf{x}$ of error probabilities is plotted as a function of the spatial position $i$ on the coupling axis for a coupled GLDPC code (see section \ref{ssection:GLDPC}) with $n=15$, $e=3$, and channel 
noise $\epsilon=0.37$. Here $L=50$ and $W=4$ (uniform window). We plot the profile at different iterations of the message-passing algorithm. The soliton traveling from left to right is plotted every 20 iterations until iteration 180.}
\label{fig:wavePropGLDPC}
\end{figure}

Under certain initial conditions and after an initial number of iterations, the profile demonstrates a \emph{solitonic} behavior. That is, after a transient phase, it appears to develop a \emph{fixed shape} that is independent of the initial condition and travels at a \emph{constant velocity} as $t$ increases.
In fact, it has been proved in \cite{KRU12} that a solitonic wave solution exists in the context of coding when transmission takes place over the Binary Erasure Channel (BEC). However, the question of the independence of the shape from the initial conditions is left open.
The soliton is illustrated in Fig. \ref{fig:wavePropGLDPC} for a Generalized LDPC (GLDPC) code (see Sec. \ref{ssection:GLDPC} for details).
%

%
\begin{figure}
\centering
\includegraphics[scale=0.27]{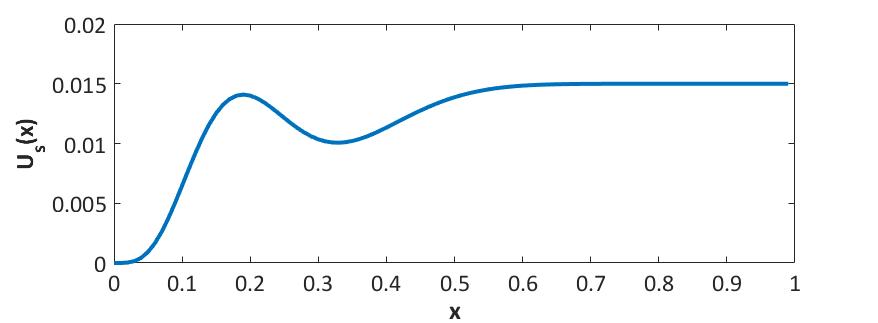}
\caption{The single potential of the GLDPC code is shown with $n=15$, $e=3$, and with channel parameter $\epsilon=0.37$. Notice that the values of the positions at which the minima occur match exactly with the boundary values of the profiles.}
\label{fig:singlePotGLDPC}
\end{figure}

In \cite{KRU12} and \cite{aref2013convergence},
bounds on the velocity of the wave for coding on the BEC are proposed. In \cite{caltagirone2014dynamics} a formula for the velocity of the wave in the context of the coupled Curie-Weiss toy model is derived and tested numerically. In \cite{elkhatibReport}, a formula in the context of coding on general binary input memoryless symmetric channels is derived.

In this work we derive a formula for the velocity of the wave in the {\it continuum limit} $L\gg w \gg 1$ for general scalar systems. 
By means of numerical simulations, we find that our formula is a very good estimate for the empirical velocity. We limit ourselves to the 
cases where the scalar recursive DE equation of the underlying uncoupled system has exactly two stable and one unstable fixed points. Equivalently the potential function (of the uncoupled system) has two minima and one local maximum.
Fig. \ref{fig:wavePropGLDPC} illustrates this setting for the GLDPC example.




\section{Preliminaries}\label{section:setup}

\subsection{Density Evolution and Potential Functions}\label{ssection:DEpotentials}
We adopt the framework and notations of \cite{yedla2014simple}.
Let $\mathcal{E}=[0,\epsilon_{\text{\tiny max}}]$, where $\epsilon_{\text{\tiny max}}\in(0,\infty)$, denote the space of parameters, and let $\mathcal{X}=[0,x_{\text{\tiny max}}(\epsilon)]$ and $\mathcal{Y}=[0,y_{\text{\tiny max}}(\epsilon)]$, such that $x_{\text{\tiny max}}(\epsilon)$, $y_{\text{\tiny max}}(\epsilon)\in(0,\infty)$ and $y_{\text{\tiny max}}(\epsilon)=g(x_{\text{\tiny max}}(\epsilon);\epsilon)$.
Consider bounded and smooth functions $f:\mathcal{Y}\times\mathcal{E}\to\mathcal{X}$ and $g:\mathcal{X}\times\mathcal{E}\to\mathcal{Y}$,  increasing in both arguments. We consider the following (uncoupled) recursion,
\begin{align}
x^{(t+1)}= f( g(x^{(t)};\epsilon);\epsilon ),\label{eqn:DEuncoupledDisc}
\end{align}
where $t\in\mathbb{N}$ denotes the iteration number. The recursion is initialized with $x^{(0)}=x_{\text{\tiny max}}$. 
Since $f(g(\mathcal{X}))\subset \mathcal{X}$, the initialization of the recursion \eqref{eqn:DEuncoupledDisc} implies that $x^{(1)}\leq x^{(0)}=x_{\text{\tiny max}}$. Moreover, since the functions $f$ and $g$ are monotonic and bounded, the recursion will converge to a limiting value $x^{(\infty)}$ when the number of iterations is large, and this limit is a fixed point since $f$ and $g$ are continuous. 

The typical picture in the context of applications such as coding or compressive sensing is as follows. It may help to think as 
$\epsilon$ as the level of noise for coding, as the inverse fraction of measurements in compressed sensing, or as the density of constraints 
in constraint-satisfaction problems. 

We define $x_{\text{\tiny good}}$ as the fixed point of \eqref{eqn:DEuncoupledDisc} obtained by the initialization $x^{(0)}=0$. We furthermore define the \emph{algorithmic threshold} $\epsilon_s$ as
\begin{align*}
\epsilon_s \triangleq \sup\{\epsilon|x^{(\infty)}=x_{\text{\tiny good}} \}.
\end{align*}
Since $f$ and $g$ are monotonic, we can see that for $\epsilon<\epsilon_s$, the recursion \eqref{eqn:DEuncoupledDisc} will converge to $x_{\text{\tiny good}}$ for any initialization of $x^{(0)}$ in $[0,x_{\text{\tiny max}}(\epsilon)]$. For $\epsilon > \epsilon_s$ new fixed points appear. {\it We will limit ourselves to systems with one extra stable fixed point} that we call $x_{\text{\tiny bad}}$ such that 
$x_{\text{\tiny bad}}> x_{\text{\tiny good}}$. For $\epsilon > \epsilon_s$ the iterations initialized at $x_{\text{\tiny max}}$ converge to $x_{\text{\tiny bad}}$. 
(It is easy to see that there also must exist an extra unstable fixed point $x_{\text{\tiny unst}}$ between these two, i.e.,
$x_{\text{\tiny good}} < x_{\text{\tiny unst}} < x_{\text{\tiny bad}}$.)

One can equivalently describe the (uncoupled) system by a {\it potential function} $U_s$ defined as
\begin{align}
U_s(x;\epsilon)=xg(x;\epsilon)-G(x;\epsilon)-F(g(x;\epsilon);\epsilon),
\end{align}
where $F(x;\epsilon)=\int_0^x\mathrm{d}s\,f(s;\epsilon)$ and $G(x;\epsilon)=\int_0^x\mathrm{d}s\,g(s;\epsilon)$. The fixed points of   \eqref{eqn:DEuncoupledDisc} can be obtained by setting the derivative with respect to $x$ of the potential $U_s(x;\epsilon)$ to zero.
The stable fixed points correspond to minima of $U_s$ and the unstable one to a local maximum of $U_s$. An example of the potential for the GLDPC code (see Sec. \ref{ssection:GLDPC}) is shown in Fig. \ref{fig:singlePotGLDPC}. There, $\epsilon > \epsilon_s$ and 
there are two minima corresponding to $x_{\text{\tiny good}} =0$ and 
$x_{\text{\tiny bad}}>0$ with one local maximum corresponding to $x_{\text{\tiny unst}}$. If we run density evolution, the iterations will get stuck at $x_{\text{\tiny bad}}>0$. In general, to check whether the analysis in this work applies to a certain application, one can plot its corresponding potential function and check that it indeed has 2 stable fixed points (minima) and 1 unstable fixed point (maximum).

To obtain the \emph{spatially coupled} system associated to the uncoupled one described above, we start by defining the ``coupling window" function that satisfies $w(z)>0$, when $0\leq z<1$, $w(z)=0$ otherwise, and $\int_\mathbb{R}\mathrm{d}z\,w(z)=1$. Then, we define the normalized function $w_W(z)=w(z)/\Big(\frac{1}{W}\sum_{j=0}^{W-1}w(\frac{j}{W})\Big)$. Remark that $\frac{1}{W}\sum_{j=0}^{W-1}w_W(\frac{j}{W})=1$ and that as $W\to\infty$, $w_W(z)\to w(z)/\int_\mathbb{R}\mathrm{d}z\,w(z)=w(z)$.

The coupled system is then obtained by taking $2L+1$ copies of the single system on the positions $i=\{-L,\dots,L\}$ and connecting them using the ``coupling matrix" $A_{j,k}=\frac{1}{W}w_W(\frac{k-j}{W})$. The discrete ``profile" $\mathbf{x}^{(t)}=\{x_{-L}^{(t)},\dots,x_{L}^{(t)}\}$ is then fixed at the boundaries as follows: $x_i^{(t)}=x_{\text{\tiny good}}$, for $i=\{-L,\dots,-L+W-1\}$ and all $t\in\mathbb{N}$, $x_i^{(t)}=x_{\text{\tiny bad}}$, for $i=\{L-W+1,\dots,L\}$ and all $t\in\mathbb{N}$. We run density evolution on the remainder of the chain. More specifically, for $i\in\{-L+W,\dots,L-W\}$, the coupled scalar recursion is
\begin{align}
x_i^{(t+1)}=\sum\limits_{j=-L}^{L}A_{j,i} \, f\Big(\sum\limits_{k=-L}^{L}A_{j,k} \, g(x_k^{(t)};\epsilon);\epsilon \Big),\label{eqn:DEcoupledDisc}
\end{align}

The boundary condition here is well adapted to study the propagation of the wave
after the transient phase is over. Indeed, simulations and heuristic arguments indicate that an initial profile that increases from a seeding value smaller or equal to $x_{\text{\tiny good}}$ (at the left boundary) to 
$x_{\text{\tiny max}}$ (on the right boundary) is attracted towards the class of profiles defined above.

The spatially coupled system can be described by a {\it potential functional} $U_c$ defined as
\begin{align}
U_c(\mathbf{x};\epsilon)=\sum_{i=-L+W}^{L-W}&\big(
x_ig(x_i;\epsilon)-G(x_i;\epsilon)\big)\nonumber\\
&-\sum_{i=-L+W}^{L-W}F\Big(\sum_{j=-L}^{L}A_{i,j}g(x_j;\epsilon);\epsilon\Big),
\end{align}
where $\mathbf{x}=\{x_{-L},\dots,x_{L}\}$. The fixed point form of Equ. \eqref{eqn:DEcoupledDisc} can be obtained by setting the derivative with respect to $\mathbf{x}$ of the potential $U_c(\mathbf{x};\epsilon)$ to zero.

A highly attractive property of spatially coupled systems is that they exhibit the so-called \emph{threshold saturation} phenomenon. That is, for all $\epsilon<\epsilon_c$ where $\epsilon_c>\epsilon_s$, the coupled recursions \eqref{eqn:DEcoupledDisc} drive the profile $\underline{x}^{(t)}=[x_{-w+1}^{(t)},\dots,x_L^{(t)}]$ to the desirable fixed point $\underline{x}^{(\infty)}=[x_{\text{\tiny good}},\dots,x_{\text{\tiny good}}]$. Here $\epsilon_c$ is a threshold defined by $U(x_{\text{\tiny good}};\epsilon_c) = U(x_{\text{\tiny bad}};\epsilon_c)$ and often called the {\it potential threshold} (note that in this equation $x_{\text{\tiny good}}$ and $x_{\text{\tiny bad}}$ themselves depend on $\epsilon_c$).

In the sequel we consider the range $\epsilon\in[\epsilon_s,\epsilon_c]$. It is for these values of the parameter $\epsilon$ that a soliton is observed. Let us repeat our basic assumption here: the recursion \eqref{eqn:DEuncoupledDisc} has exactly two {\it stable} fixed points $x_{\text{\tiny good}}$ and $x_{\text{\tiny bad}}$. With more than two stable fixed points the propagating wave has a more complicated structure
and our formulas would have to be adapted accordingly (see \cite{aref2013convergence} for a nice discussion  of this issue in coding).

\subsection{Continuum Limit}
We consider the system in the {\it continuum limit}, which is obtained by first taking $L\to\infty$ and then $W\to\infty$ \cite{KRU12}, \cite{el2013displacement}, \cite{el2014analysis}.
We set $\mathtt{x}(\frac{i}{W},t)\equiv \mathtt{x}_i^{(t)}$ and replace $\frac{i}{W}\to z$, $\frac{j}{W}\to u$, $\frac{k}{W}\to s$, 
where $z$, $u$, $s\in\mathbb{R}$ are {\it continuous} spatial variables. The coupled recursion, in the continuum limit, can then be written as
\begin{align}
\resizebox{.91\hsize}{!}{$x(z,t+1)=\int_0^1\mathrm{d}u\,w(u)\, f\Big(\int_0^1\mathrm{d}s\,w(s)\, g(x(z-u+s,t);\epsilon);\epsilon \Big).$}\label{eqn:DEcoupledCont}
\end{align}
The boundary conditions on the continuous profile $x(\cdot,\cdot)$ become $x(z,t)\to x_{\text{\tiny good}}$ when $z\to -\infty$ 
and $x(z,t)\to x_{\text{\tiny bad}}$ when $z\to+\infty$. Again, this boundary condition captures the profiles obtained after the transient phase has passed, and is well adapted to the study of the wave propagation.

Let $x_0(z)$ be a static (time independent) profile that satisfies the boundary conditions $x_0(z)\to x_{\text{\tiny good}}$ when $z\to -\infty$ 
and $x_0(z)\to x_{\text{\tiny bad}}$ when $z\to+\infty$. This profile can be thought as an initial condition for the recursions. For us however it serves as a reference profile in order 
to define the potential functional in the continuum limit. We look at the continuous version of $U(\mathbf{x};\epsilon)$ which we call $\mathcal{W}[x(\cdot,\cdot);\epsilon]$ and subtract from it $\mathcal{W}[x_0(\cdot);\epsilon]$ so that the integrals converge. The potential functional $\Delta\mathcal{W}[x(\cdot,\cdot);\epsilon]$ in the continuum limit is thus {\it defined} as
\begin{align*}
&\Delta\mathcal{W}[x(\cdot,\cdot);\epsilon]\triangleq\int_\mathbb{R}\mathrm{d}z\,\Big\{x(z,t)g(x(z,t);\epsilon)-x_0(z)g(x_0(z);\epsilon)\\
&\qquad-G(x(z,t);\epsilon)-F\Big(\int_0^1\mathrm{d}u\,w(u)\,g(x(z-u,t);\epsilon);\epsilon\Big)\\
&\qquad+G(x_0(z);\epsilon)+F\Big(\int_0^1\mathrm{d}u\,w(u)\,g(x_0(z-u);\epsilon);\epsilon \Big)\Big\}.
\end{align*}
As long as $x_0(z)$ converges to its limiting values fast enough the integrals over the spatial axis are well defined.
We can further split the continuous potential functional into two parts: the single potential $\mathcal{W}_s[x(\cdot,\cdot);\epsilon]$ (that we obtain setting $w(z) \to 0$) and the interaction potential $\mathcal{W}_i[x(\cdot,\cdot);\epsilon]$ (that is caused only by coupling), defined as follows
\begin{align*}
\mathcal{W}_s[x(\cdot,\cdot);\epsilon]\triangleq\int_\mathbb{R}\mathrm{d}z\,\Big\{& x(z,t)g(x(z,t);\epsilon)-x_0(z)g(x_0(z);\epsilon)\\
&-G(x(z,t);\epsilon)-F(g(x(z,t);\epsilon);\epsilon)\\
&+G(x_0(z);\epsilon)-F(g(x_0(z);\epsilon);\epsilon)\Big\},\\
\mathcal{W}_i[x(\cdot,\cdot);\epsilon]\triangleq\int_\mathbb{R}\mathrm{d}z\,\Big\{& F(g(x(z,t);\epsilon);\epsilon)+F(g(x_0(z);\epsilon);\epsilon)\\
&-F\Big(\int_0^1\mathrm{d}u\,w(u)g(x(z-u,t);\epsilon);\epsilon\Big)\\
&+F\Big(\int_0^1\mathrm{d}u\,w(u)\,g(x_0(z-u);\epsilon) ;\epsilon\Big)\Big\}.
\end{align*}
\section{Velocity for General Scalar Systems}\label{section:formula}
\subsection{Statement of main result}
We assume that after a number of iterations, which we call the transient phase, the density profile $x(\cdot,\cdot)$ develops a {\it fixed shape} which we call $X(\cdot)$ that moves with {\it constant velocity} $v$. That is, we make the ansatz $x(z,t)=X(z-vt)$.

Then, we find the following formula for the velocity $v$
\begin{align}\label{eqn:vFormula}
v=\frac{U_s(x_{\text{\tiny bad}};\epsilon)-U_s(x_{\text{\tiny good}};\epsilon)}{\int_{\mathbb{R}}\mathrm{d}z\,g'(X(z);\epsilon)(X'(z))^2},
\end{align}
where $g^\prime=\partial_x g$ is the derivative of $g$ with respect to its first argument, and $X^\prime$ is the derivative of the profile.

\subsection{Derivation of main result}\label{ssection:mainFormulaDeriv}
Evaluating the functional derivative of $\Delta\mathcal{W}[x(\cdot,\cdot);\epsilon]$ in an arbitrary direction $\eta(\cdot,\cdot)$, and then using \eqref{eqn:DEcoupledCont} we obtain
\begin{align*}
&\frac{\delta\Delta\mathcal{W}[x(\cdot,\cdot);\epsilon]}{\delta x(\cdot,\cdot)}[\eta(z,t)]\\
&=\lim_{\gamma\to 0}\frac{1}{\gamma}\Big\{\Delta\mathcal{W}[x(\cdot,\cdot)+\gamma\eta(\cdot,\cdot);\epsilon]-\Delta\mathcal{W}[x(\cdot,\cdot);\epsilon]\Big\}\\
&=\int_\mathbb{R}\mathrm{d}z\,\eta(z,t)g'(x(z,t);\epsilon)\Big\{x(z,t)\\
&\qquad\quad-\int_0^1{d}u\,w(u)f(\int_0^1{d}s\,w(s)g(x(z-u+s,t);\epsilon);\epsilon) \Big\}\\
&=-\int_\mathbb{R}\mathrm{d}z\,\eta(z,t)g'(x(z,t);\epsilon)\big(x(z,t+1)-x(z,t)\big)
\end{align*}
Using the ansatz $x(z,t)=X(z-vt)$ and the approximation $x(z,t+1)-x(z,t)\approx -vX'(z)$, the functional derivative of $\Delta\mathcal{W}[x(\cdot,\cdot);\epsilon]$ in the special direction $\eta(z,t)=X'(z)$ becomes
\begin{align*}
\frac{\delta\Delta\mathcal{W}[x(\cdot,\cdot);\epsilon]}{\delta x(\cdot,\cdot)}&[X'(z)]=v\int_\mathbb{R}\mathrm{d}z\,(X'(z))^2g'(X(z);\epsilon).
\end{align*}
To obtain our formula \eqref{eqn:vFormula} we separate the left-hand side of the equation above 
into two contributions
\begin{align*}
\frac{\delta\mathcal{W}_s[X(\cdot);\epsilon]}{\delta X(\cdot)}[X'(z)]+\frac{\delta\mathcal{W}_i[X(\cdot);\epsilon]}{\delta X(\cdot)}[X'(z)]
\end{align*}
and calculate each term separately. For the first (uncoupled) part we find
\begin{align*}
&\frac{\delta\mathcal{W}_s[X(\cdot);\epsilon]}{\delta X(\cdot)}[X'(z)]\\
&=\int_\mathbb{R}\mathrm{d}z\,X'(z)\big(X(z)g'(X(z);\epsilon)-g'(X(z);\epsilon)f(g(X(z);\epsilon);\epsilon)\big)\\
&=\int_\mathbb{R}\mathrm{d}z\,\frac{\mathrm{d}}{\mathrm{d}z}\big\{X(z)g(X(z);\epsilon)-G(X(z);\epsilon)-F(g(X(z);\epsilon);\epsilon)\big\}\\
&=\Bigl[U_s(X(z);\epsilon)\Bigr]_{-\infty}^{+\infty}=U_s(x_{\text{\tiny bad}};\epsilon)-U_s(x_{\text{\tiny good}};\epsilon).
\end{align*}
For the second (interaction) part we find
\begin{align*}
&\frac{\delta\mathcal{W}_i[X(\cdot);\epsilon]}{\delta X(\cdot)}[X'(z)]\\
&=\int_\mathbb{R}\mathrm{d}z\frac{\mathrm{d}}{\mathrm{d}z}\Big\{F(g(X(z);\epsilon);\epsilon)\\
&\qquad\qquad\qquad -F\big(\int_0^1\mathrm{d}u\,w(u)g(X(z-u);\epsilon);\epsilon\big)\Big\}\\
&=\Bigl[F(g(X(z);\epsilon);\epsilon)-F\big(\int_0^1\mathrm{d}uw(u)g(X(z-u);\epsilon);\epsilon\big)\Bigr]_{-\infty}^{+\infty}=0
\end{align*}

\section{Applications}\label{section:applications}
The general formula for the velocity of the soliton for scalar systems \eqref{eqn:vFormula} can be applied on several examples. In particular we recover the results of \cite{elkhatibReport} for standard LDPC codes over the BEC, as well as on general Binary Memoryless Symmetric (BMS) channels
within the scalar Gaussian approximation. 
In this section we provide two more scalar applications, namely GLDPC codes and compressive sensing.

The predictions of our formula are compared with the observed, empirical velocity $v_e$ that is obtained by running the scalar recursions. To obtain $v_e$, we plot the discrete profile $\mathbf{x}$ at different iterations of the scalar recursions and find the average of $\Delta z/(W\Delta I)$, where $\Delta z$ is the spatial difference between the kinks of the profiles, $\Delta I$ is the difference in the number of iterations, and $W$ is the size of the coupling window. 

\subsection{Generalized LDPC Codes}\label{ssection:GLDPC}
We consider a GLDPC code described as follows: The variable (bit) nodes have degree 2 and the check nodes have degree $n$. The rules of the check nodes are given by a primitive BCH code of blocklength $n$ (unlike LDPC codes where they are parity checks). An attractive property of BCH codes is that they can be designed to correct a chosen number of errors. For instance, one can design a BCH code so that it corrects all patterns of at most $e$ erasures on the BEC, and all error patterns of weight at most $e$ on the Binary Symmetric Channel (BSC).


We consider transmission on the BEC or BSC and denote by $\epsilon$ the channel parameter. The density evolution recursions have been derived for both channels, based on a bounded distance decoder for the BCH code 
\cite{Henry-Narayan-et-al}.
We have $\epsilon_{\text{\tiny max}}=x_{\text{\tiny max}}=y_{\text{\tiny max}}=1$, and for $n$ and $e$ fixed \cite{yedla2014simple},
\begin{align*}
f(x;\epsilon)&=\epsilon x,\\
g(x,e,n;\epsilon)&=\sum_{i=e}^{n-1}{{n-1}\choose{i}}x^i(1-x)^{n-i-1}.
\end{align*}

%
The formula for the velocity of the soliton appearing in the case of the GLDPC codes is found from \eqref{eqn:vFormula},
where the single potential of the system $U_{\text{\tiny GLDPC}}(\cdot)$ is given by
\begin{align*}
U_{\text{\tiny GLDPC}}(x,e,n;\epsilon)=\frac{e}{n}g(x,e,n;\epsilon)-&\frac{x(1-x)}{n}g'(x,e,n;\epsilon)\\
&-\frac{\epsilon}{2}g^2(x,e,n;\epsilon).
\end{align*}
\begin{figure}
\centering
\includegraphics[scale=0.27]{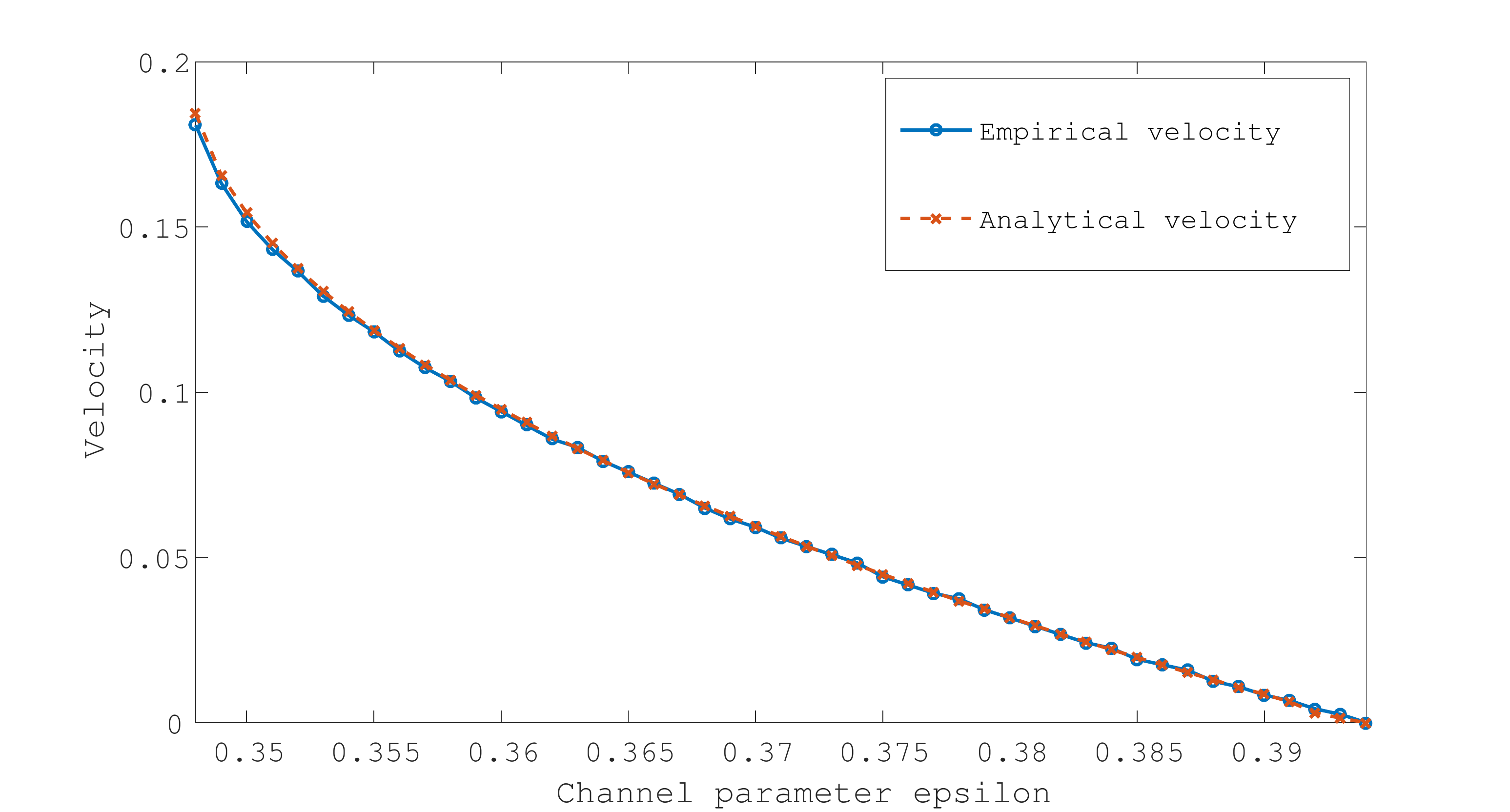}
\caption{We consider a GLDPC code with $n=15$ and $e=3$, with spatial length $L=250$ and uniform coupling window with $W=3$. We plot the velocities (normalized by $W$) $v_{\text{\tiny GLDPC}}$ and $v_e$ as a function of the channel parameter $\epsilon$ when $\epsilon$ is between the BP threshold $\epsilon_s=\epsilon_{\text{\tiny BP}}= 0.348$ and the potential threshold $\epsilon_c=\epsilon_{\text{\tiny MAP}}\approx 0.394$.}
\label{fig:velocitiesGLDPC}
\end{figure}

\begin{figure}
\centering
\includegraphics[scale=0.27]{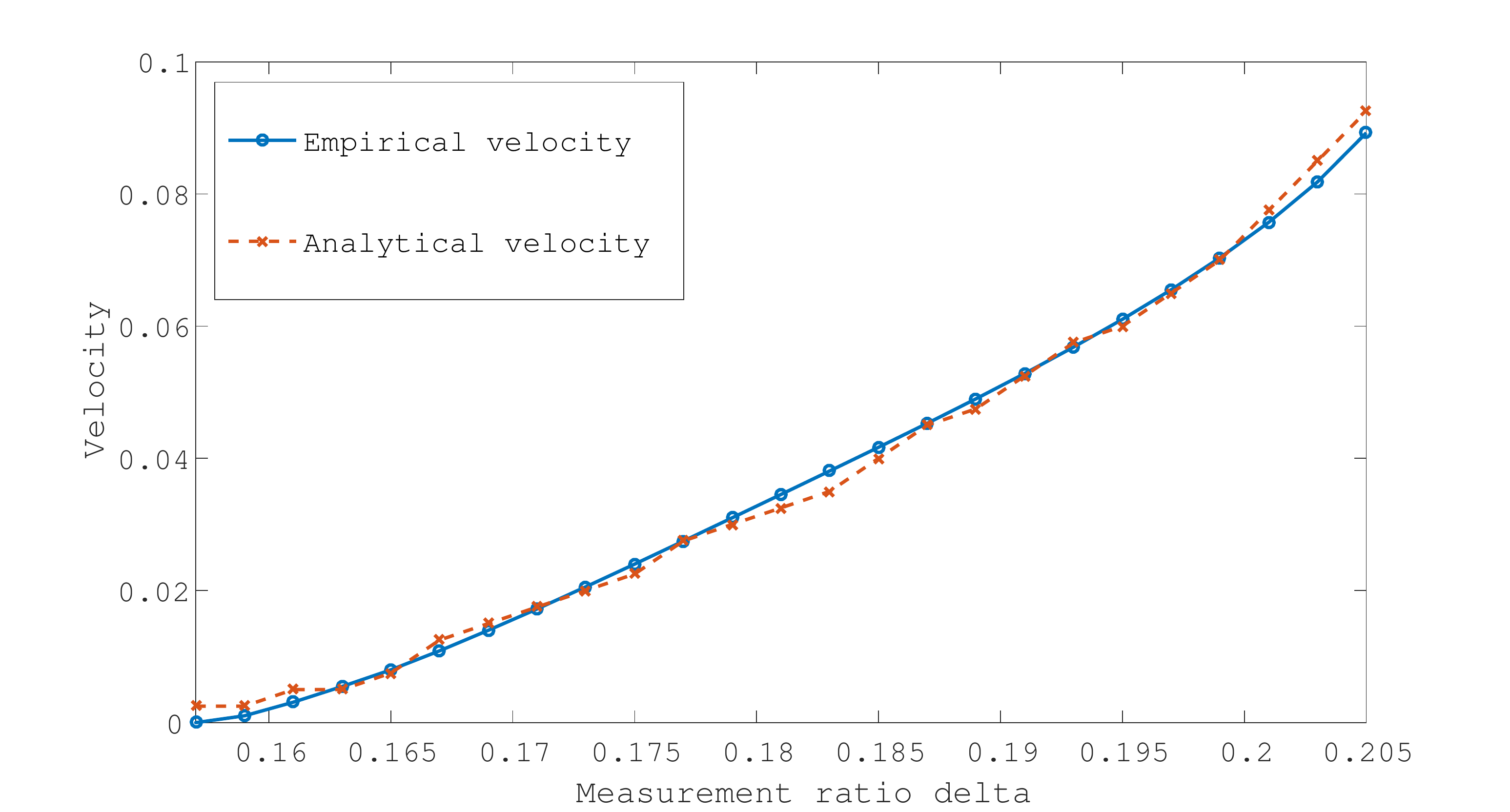}
\caption{We consider the compressive sensing problem with $\mathtt{snr}=10^5$ and Gaussian-Bernoulli prior for the signal components with sparsity parameter $\rho=0.1$. We have $L=250$ and uniform coupling window with $W=4$. We plot the velocities (normalized by $W$) $v_{\text{\tiny CS}}$ and $v_e$ as a function of the measurement fraction $\delta$ when $\delta$ is between the potential threshold $\delta_c=0.157$ and $\delta_s= 0.208$.}
\label{fig:velocitiesCS}
\end{figure}

\begin{figure}
\centering
\includegraphics[scale=0.27]{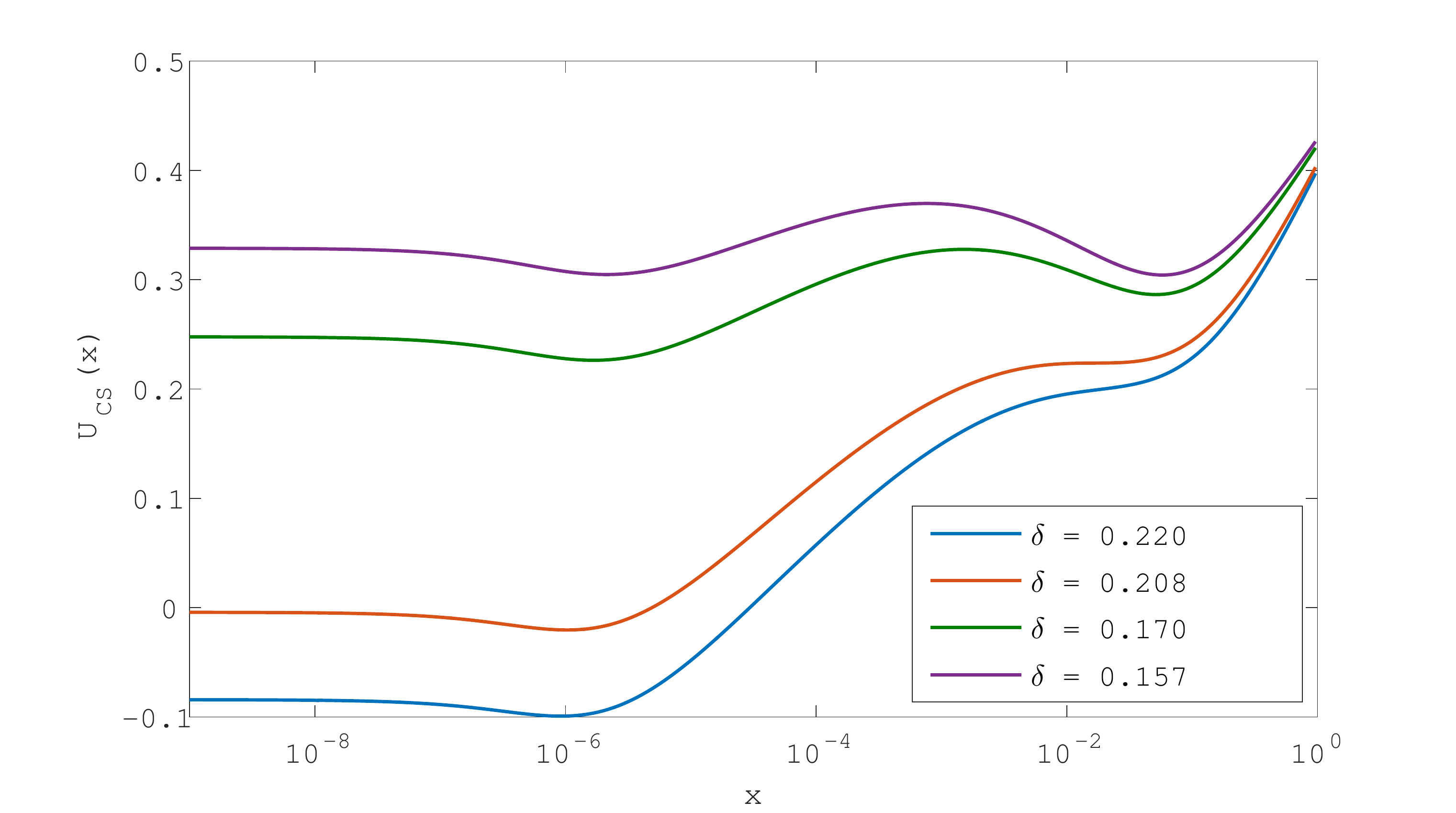}
\caption{We consider the compressive sensing problem with $\mathtt{snr}=10^5$ and Gaussian-Bernoulli prior for the signal components with sparsity parameter $\rho=0.1$. We plot the single potential for several values of $\delta\in[\delta_c,\delta_s]=[0.157,0.208]$. We can see that when $\delta > \delta_s$ the potential has a unique minimum, when $\delta=\delta_s$, there appears an inflection point, for $\delta_c<\delta<delta_s$ the potential has two minima and the energy gap $\Delta_E$ is strictly positive, and for $\delta=\delta_c$ the energy gap vanishes.}
\label{fig:velocitiesCS}
\end{figure}

Figure \ref{fig:velocitiesGLDPC} shows the velocities (normalized by $W$) for the spatially coupled GLPDC code with $n=15$ and $e=3$, when the coupling parameters satisfy $L=250$ and $W=3$ and we use the uniform coupling window. We plot the velocities for $\epsilon\in[\epsilon_s,\epsilon_c]=[0.348,0.394]$. We observe that the formula for the velocity provides a very good estimation of the empirical velocity $v_e$.

\subsection{Compressive Sensing}
Let $\mathbf{s}$ be a length-$n$ signal vector (where the components are i.i.d. copies of a random variable $S$) which is acquired 
through $m$ linear measurements. We assume that the measurement matrix has i.i.d Gaussian elements $\mathcal(0, 1/\sqrt n)$. 
We call $\delta =m /n$ the fixed measurement ratio when $n\to\infty$. The relation between $\delta$ and the generic $\epsilon$ used in this paper
is $\epsilon = \delta^{-1}$. 
We assume that $\mathbb{E}[S^2]=1$ and that each component of $\mathbf{s}$ is corrupted by independent Gaussian noise of variance $\sigma^2=1/\mathtt{snr}$.
To recover $\mathbf{s}$ one implements the so-called approximate message-passing (AMP) algorithm. It has been shown that the analysis of this algorithm is given by state evolution \cite{donoho2012information}. 

We consider the approach described in \cite{yedla2014simple}. Consider the minimum mean-squared error (mmse) function given by 
$\mathtt{mmse}(\mathtt{snr})=\mathbb{E}_{S,Y}[(S-\hat S(Y,\mathtt{snr}))^2]$ where $\hat S(Y,\mathtt{snr})=\mathbb{E}_{S|Y}[S|Y]$
and $Y=\sqrt{\mathtt{snr}}S+Z$, $Z\sim \mathcal{N}(0,1)$. The state evolution equations, which track the mean squared error of the AMP estimate, then correspond to the recursion defined by 
\begin{align*}
f(x,\mathtt{snr};\delta)&=\mathtt{mmse}(\mathtt{snr}-x),\\
g(x,\mathtt{snr};\delta)&=\mathtt{snr}-\frac{1}{\frac{1}{\mathtt{snr}}+\frac{x}{\delta}},
\end{align*}
where $\delta_{\text{\tiny max}}=1$, $x_{\text{\tiny max}}=\mathtt{mmse}(0)$, $y_{\text{\tiny max}}=g(x_{\text{\tiny max}})$.

The formula for the velocity $v_{\text{\tiny CS}}$ of the soliton in the case of compressive sensing can be obtained from \eqref{eqn:vFormula} 
where the single potential $U_{\text{\tiny SC}}(\cdot)$ of the system is given by
\begin{align*}
U_{\text{\tiny SC}}(x;\delta)=-&\frac{x}{\frac{1}{\mathtt{snr}}+\frac{x}{\delta}}+\delta\ln\Big(1+\frac{x}{\delta/\mathtt{snr}}\Big)\\
&-2I\Big(S;\sqrt{\mathtt{snr}}S+Z\Big)+2I\Big(S;\sqrt{\frac{1}{\frac{1}{\mathtt{snr}}+\frac{x}{\delta}}}S+Z\Big).
\end{align*}
To check that this is indeed the correct potential we differentiate $U_{\text{\tiny SC}}(x;\epsilon)$ with respect to $x$  and use the relation \cite{guo2005mutual} $\frac{1}{2}\mathtt{mmse}(\mathtt{snr})=\frac{\mathrm{d}}{\mathrm{d}\mathtt{snr}}I(S;\sqrt{\mathtt{snr}}S+Z)$, where $I(A;B)$ denotes the mutual information between the random variables $A$ and $B$ and is measured in nats.

For the compressive sensing scheme, we assume that the prior distribution on $S$ is the Bernoulli-Gaussian described by
\begin{align*}
q_0(y)=(1-\rho)\delta(y)+\rho\Phi_0(y),
\end{align*}
where $\Phi_0(y)=(1/\sqrt{2\pi})e^{-y^2/2}$. Figure \ref{fig:velocitiesCS} shows the velocities (normalized by $W$) for the spatially coupled compressive sensing scheme with $\mathtt{snr}=10^5$ and $\rho=0.1$, when the coupling parameters satisfy $L=250$ and $W=4$ and we use the uniform coupling window. We remark that for this application, the potential threshold $\delta_c$ is smaller than $\delta_s$ because the smaller the value of $\delta$, the less measurements of the signal we make, which induces more uncertainty. We plot on Fig. \ref{fig:velocitiesCS} the velocities for $\delta\in[\delta_c,\delta_s]=[0.157,0.208]$. Similarly, we observe that the formula for the velocity provides a good estimation of the empirical velocity $v_e$. 

\section*{Acknowledgments}
R. E. thanks Tongxin Li for fruitful discussions and Jean Barbier for giving her part of his code on compressive sensing and for patiently explaining it to her. Part of this work was done during the authors' stay at the Institut Henri Poincar\'e - Centre \'Emile Borel. The authors thank this institution for its hospitality and support.

\bibliographystyle{IEEEtran}
\bibliography{BIBfile}

\end{document}